\documentclass[iop]{emulateapj}  

\usepackage{color}

\newcommand\kms{\ifmmode{\rm km\thinspace s^{-1}}\else km\thinspace s$^{-1}$\fi}
\newcommand\epic{EPIC~219568666}
\newcommand\epicfirst{EPIC~219394517}
\newcommand\ktwo{{\it K2\/}}
\newcommand\kepler{{\it Kepler\/}}

\shortauthors{Torres et al.}
\shorttitle{Eclipsing binary in Ruprecht\thinspace 147}

\begin{document} 
\submitted{Accepted for publication in The Astrophysical Journal}

\title{Eclipsing binaries in the open cluster Ruprecht\thinspace 147. II: \epic}

\author{
Guillermo Torres\altaffilmark{1},
Andrew Vanderburg\altaffilmark{2,3},
Jason L.\ Curtis\altaffilmark{4},
David Ciardi\altaffilmark{5},
Adam L.\ Kraus\altaffilmark{2},
Aaron C.\ Rizzuto\altaffilmark{2},
Michael J. Ireland\altaffilmark{6},
Michael B.\ Lund\altaffilmark{5},
Jessie L.\ Christiansen\altaffilmark{5}, and
Charles A.\ Beichman\altaffilmark{5}
}

\altaffiltext{1}{Center for Astrophysics $\vert$ Harvard \&
  Smithsonian, 60 Garden St., Cambridge, MA 02138, USA;
  gtorres@cfa.harvard.edu}

\altaffiltext{2}{Department of Astronomy, The University of Texas at
  Austin, Austin, TX 78712, USA}

\altaffiltext{3}{NASA Sagan Fellow}

\altaffiltext{4}{Department of Astronomy, Columbia University, New
  York, NY 10027, USA}

\altaffiltext{5}{Caltech/IPAC-NASA Exoplanet Science Institute,
  Pasadena, CA 91125, USA}

\altaffiltext{6}{Research School of Astronomy and Astrophysics,
  Australian National University, Canberra, ACT 2611, Australia}

\begin{abstract}

We report our spectroscopic monitoring of the detached, grazing, and
slightly eccentric 12-day double-lined eclipsing binary \epic\ in the
old nearby open cluster Ruprecht\thinspace 147. This is the second
eclipsing system to be analyzed in this cluster, following our earlier
study of \epicfirst.  Our analysis of the radial velocities combined
with the light curve from the \ktwo\ mission yield absolute masses and
radii for \epic\ of $M_1 = 1.121 \pm 0.013~M_{\sun}$ and $R_1 = 1.1779
\pm 0.0070~R_{\sun}$ for the F8 primary, and $M_2 = 0.7334 \pm
0.0050~M_{\sun}$ and $R_2 = 0.640 \pm 0.017~R_{\sun}$ for the faint
secondary. Comparison with current stellar evolution models calculated
for the known metallicity of the cluster points to a primary star that
is oversized, as is often seen in active M dwarfs, but this seems
rather unlikely for a star of its mass and with a low level of
activity.  Instead, we suspect a subtle bias in the radius ratio
inferred from the photometry, despite our best efforts to avoid it,
which may be related to the presence of spots on one or both stars.
The radius sum for the binary, which bypasses this possible problem,
indicates an age of $2.76 \pm 0.61$~Gyr that is in good agreement with
a similar estimate from the binary in our earlier study.

\end{abstract}


\section{Introduction}
\label{sec:introduction}

Detached, double-lined eclipsing binaries are the primary source of
precise empirical masses and radii for normal stars \citep[see,
  e.g.,][]{Andersen:1991, Torresetal:2010}. When coupled with stellar
evolution models they can also provide accurate ages. If the binary
happens to be located in a cluster, the age determination serves as an
important check against more traditional estimates obtained by
isochrone fitting or gyrochronology, or with other techniques of
dating single member stars such as asteroseismology.

Ruprecht\thinspace 147 (NGC~6774) is a nearby, middle-aged open
cluster with a slightly supersolar chemical composition
\citep[distance $\sim$300~pc, age $\sim$3~Gyr, ${\rm [Fe/H]} =
  +0.10$;][]{Curtis:2013}. It is endowed with no fewer than 5 detached
eclipsing binaries identified by \cite{Curtis:2016} that are
relatively bright and amenable to detailed investigation. In an
earlier study we reported results for the first of these,
\epicfirst\ \citep[][Paper~I]{Torres:2018}, a pair of very similar
early G-type stars in a 6.5-day orbit that yielded highly accurate
masses and radii good to better than 0.2\% and 1\%,
respectively. These properties gave an excellent fit to current
stellar evolution models, and an improved age for the cluster between
about 2.5 and 2.6~Gyr, depending on the model, with a formal
uncertainty of about 0.3~Gyr.

The present paper reports on our follow-up observations and analysis
of a second eclipsing binary system in Ruprecht\thinspace 147,
\epic\ (TYC 6296-2012-1, 2MASS J19165992$-$1625176, Gaia DR2
4183920989590558720, $V = 11.86$), which has the potential to add
significantly to the characterization of the cluster. By virtue of the
location of Ruprecht\thinspace 147 on the ecliptic, \epic\ was
observed along with the other eclipsing systems by NASA's
\ktwo\ mission in the final months of 2015 (Campaign 7). Initial
follow-up by \cite{Curtis:2016} determined it to be a 12-day period,
slightly eccentric single-lined spectroscopic binary, which we find
here to be double-lined. However, unlike the previous binary we
reported on, the mass ratio in this case is very different from unity,
which makes the secondary very faint.  Membership in the cluster was
established by \cite{Curtis:2016}, and confirmed more recently based
on parallax and proper motion information from the {\it Gaia\/}
mission by \cite{Cantat-Gaudin:2018}.

As we describe below, the seemingly straightforward analysis of the
high-precision \ktwo\ light curve of \epic\ turns out to present
difficulties that compromise the accuracy of the results. Although
this does not allow us to take full advantage of the precise
individual mass and radius measurements to test models of stellar
evolution, the system still yields a useful estimate of the age of the
cluster. This case serves as an interesting illustration of the
pernicious effects of systematic errors that can easily go unnoticed,
particularly those in the photometry that can bias the determination
of the stellar radii in a significant way.


The layout of the paper is as follows. Section~\ref{sec:observations}
describes our reduction and treatment of the raw \ktwo\ photometry of
\epic, our new high-resolution spectroscopic observations, and
archival and new imaging observations to explore the field surrounding
the binary. In Section~\ref{sec:analysis} we present the joint
spectroscopic and light curve analysis, leading to the physical
properties of the system discussed in Section~\ref{sec:dimensions}. We
compare these properties with current stellar evolution models as well
as with the properties of the binary in our earlier study in
Section~\ref{sec:models}. While this reveals discrepancies that could
be due to a bias in the radius ratio preventing us from relying on the
individual radii to infer an age, we are still able to make use of the
radius sum, which appears to be well determined.
Section~\ref{sec:discussion} then discusses multiple tests to
investigate the source of the discrepancies.
%
%
Closing remarks are given in Section~\ref{sec:remarks}.

\section{Observations}
\label{sec:observations}

\subsection{Photometry}
\label{sec:photometry}

\epic\ was observed by \ktwo\ during its 7th observing campaign for 81
days between October and December of 2015, with a cadence of 29.4
minutes (3654 measurements). The target fell within a large
super-aperture designed to observe many members of the
Ruprecht\thinspace 147 cluster together.  Following downlink from the
spacecraft and calibration by the NASA Ames \ktwo\ pipeline, we
proceeded to download the pixel time series from the
Ruprecht\thinspace 147 super-aperture from the Mikulski Archive for
Space Telescopes (MAST).\footnote{\url{http://archive.stsci.edu/}}
Because \epic\ is located in a dense region of stars, we extracted the
light curve by calculating the flux within a concentric set of 10
circular moving apertures, to ensure that third-light contamination
from nearby stars was kept constant as the telescope's pointing
drifted. We performed a first-pass correction for \ktwo's spacecraft
systematics using the methods described by \cite{Vanderburg:2014} and
\cite{Vanderburg:2016}, and selected the circular moving aperture that
yielded the highest photometric precision after the systematics
correction. The aperture we selected had a radius of 3.97 pixels, or
15\farcs80. We then refined the systematics correction by
simultaneously fitting for the eclipse shapes using \cite{Mandel:2002}
transit models, \ktwo\ roll systematics using splines as a function of
\kepler's pointing position, and long-term variability using splines
in time, as described by \cite{Vanderburg:2016}. The resulting light
curve has a scatter of about 100~ppm per 30 minute exposure, and
contains 7 primary eclipses and 6 secondary eclipses. We removed
low-frequency variability by dividing away the best-fit spline from
our simultaneous light curve solution. The data processed in this way
(3654 measurements) are given in Table~\ref{tab:photometry}.

\begin{deluxetable}{cc}
\tablewidth{0pt}
\tablecaption{Detrended \ktwo\ Photometry of \epic\ \label{tab:photometry}}
\tablehead{
\colhead{HJD} &
\colhead{}
\\
\colhead{(2,400,000+)} &
\colhead{Residual flux}
}
\startdata
     57301.4866   &    0.99987691 \\
     57301.5070   &    0.99997035 \\
     57301.5275   &    1.00011420 \\
     57301.5479   &    1.00008800 \\
     57301.5683   &    1.00009520
\enddata
\tablecomments{(This table is available in its entirety in machine-readable form.)}
\end{deluxetable}

\subsection{Imaging}
\label{sec:imaging}

\begin{figure}
\epsscale{1.15}
\plotone{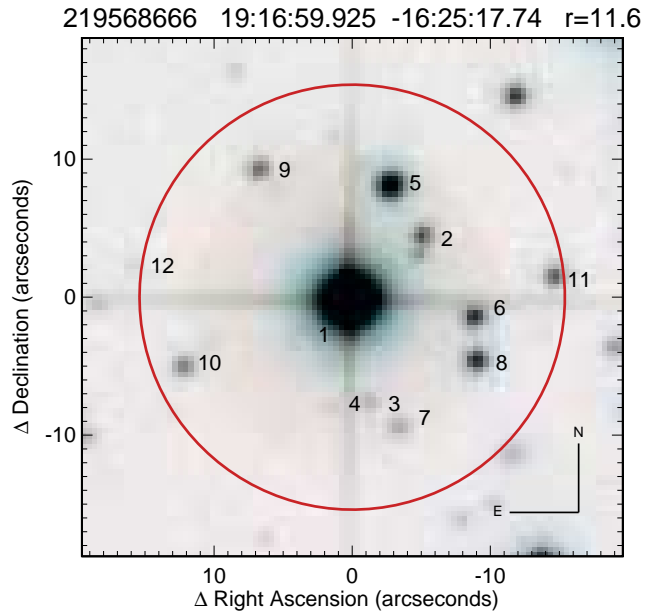}
\figcaption{CFHT $r$-band image of the field of \epic, with the
  15\farcs80 photometric aperture used to extract the \ktwo\ photometry
  indicated with a circle. Nearby companions are numbered as in
  Table~\ref{tab:CFHT}. Companions \#1 through \#4 are too faint to
  influence the light curve analysis.\label{fig:CFHT}}
\end{figure}

Several fainter stars near \epic\ fall within the 15\farcs80 aperture
we used to extract the photometry, and can potentially affect the
parameters derived from the light curve. In Figure~\ref{fig:CFHT} we
show an image of the target in a passband similar to Sloan $r$ taken
in 2008 by \cite{Curtis:2013} with the MegaCam instrument
\citep{Hora:1994} on the Canada-France-Hawaii Telescope (CFHT). We
measured the position relative to the target and the brightness in the
CFHT $gri$
filters\footnote{\url{http://www.cadc-ccda.hia-iha.nrc-cnrc.gc.ca/en/megapipe/docs/filtold.html}}
of all numbered companions within the aperture, and list them in
Table~\ref{tab:CFHT}. This information is used later in
Section~\ref{sec:analysis} to make a quantitative estimate of the flux
contamination. The table includes $J$- and $K$-band brightness
measurements based on UKIRT/WFCAM imaging \citep{Curtis:2016}. We also
report the properties of several of these companions that are listed
in the {\it Gaia}/DR2 catalog \citep{Gaia:2018}, though none appear to
be members of the cluster, based on their parallax.


\setlength{\tabcolsep}{4.5pt}
\begin{deluxetable*}{rccccccccccccc}
\tablewidth{0pt}
\tablecaption{Neighbors of \epic\ \label{tab:CFHT}}
\tablehead{
\colhead{} &
\colhead{R.A.} &
\colhead{Dec.} &
\colhead{P.A.} &
\colhead{$\rho$} & 
\colhead{$g$} &
\colhead{$r$} &
\colhead{$i$} &
\colhead{$\sigma(gri)$} & 
\colhead{$J$} &
\colhead{$K$} &
\colhead{$\sigma(JK)$} &
\colhead{$G$} &
\colhead{$\pi_{\rm DR2}$}
\\
\colhead{\#} &
\colhead{(J2000)} &
\colhead{(J2000)} &
\colhead{(degree)} &
\colhead{($\arcsec$)} &
\colhead{(mag)} &
\colhead{(mag)} &
\colhead{(mag)} &
\colhead{(mag)} &
\colhead{(mag)} &
\colhead{(mag)} &
\colhead{(mag)} &
\colhead{(mag)} &
\colhead{(mas)}
}
\startdata
 1 & 19:17:00.172 & $-$16:25:20.98 & 131.8    & \phn4.9 &  \nodata  &  \nodata  &  \nodata  &  \nodata  & 18.38 & 17.58 & 0.05 &  \nodata  &  \nodata          \\
 2 & 19:16:59.577 & $-$16:25:13.18 & 312.7    & \phn7.0 &  \nodata  &  \nodata  &  \nodata  &  \nodata  & 17.03 & 16.62 & 0.02 & 18.59     & $0.398 \pm 0.344$ \\
 3 & 19:16:59.832 & $-$16:25:25.20 & 190.4    & \phn7.5 &  \nodata  &  \nodata  &  \nodata  &  \nodata  & 17.79 & 16.95 & 0.03 & 20.62     &  \nodata          \\
 4 & 19:17:00.023 & $-$16:25:25.56 & 169.8    & \phn7.9 &  \nodata  &  \nodata  &  \nodata  &  \nodata  & 18.78 & 17.71 & 0.05 &  \nodata  &  \nodata          \\
 5 & 19:16:59.731 & $-$16:25:09.50 & 341.3    & \phn8.8 & 16.65     & 16.12     & 15.93     & 0.02      & 14.99 & 14.55 & 0.02 & 16.36     & $0.348 \pm 0.077$ \\
 6 & 19:16:59.323 & $-$16:25:18.97 & 262.4    & \phn9.1 & 17.67     & 17.26     & 17.07     & 0.02      & 16.68 & 16.25 & 0.02 & 17.98     & $0.071 \pm 0.223$ \\
 7 & 19:16:59.686 & $-$16:25:26.91 & 200.9    & \phn9.8 & 21.09     & 20.64     & 20.20     & 0.11      & 18.45 & 17.84 & 0.06 & 20.27     & $0.850 \pm 1.229$ \\
 8 & 19:16:59.313 & $-$16:25:22.11 & 244.1    & 10.2    & 18.43     & 17.82     & 17.54     & 0.02      & 16.23 & 15.60 & 0.02 & 17.97     & $0.401 \pm 0.184$ \\
 9 & 19:17:00.361 & $-$16:25:08.31 & \phn33.4 & 11.5    & 18.72     & 18.38     & 18.29     & 0.02      & 17.54 & 17.07 & 0.03 & 18.89     & $0.021 \pm 0.311$ \\
10 & 19:17:00.724 & $-$16:25:22.56 & 112.4    & 12.9    & 20.06     & 19.43     & 19.17     & 0.05      & 17.77 & 17.31 & 0.04 & 19.31     & $1.280 \pm 0.455$ \\
11 & 19:16:58.939 & $-$16:25:16.02 & 277.2    & 14.9    & 19.60     & 18.88     & 18.63     & 0.04      & 17.25 & 16.70 & 0.02 & 18.82     & $0.293 \pm 0.317$ \\
12 & 19:17:00.952 & $-$16:25:15.29 & \phn80.3 & 15.6    & 25.82     & 24.04     & 22.62     & 0.81      & 18.61 & 17.65 & 0.05 &  \nodata  &  \nodata          
\enddata
\tablecomments{Coordinates derived from the astrometric solutions of the CFHT images \citep[see][]{Curtis:2013}.}
\end{deluxetable*}

In order to search for blended stellar companions to \epic\ inside the
inner working angle of the CFHT imaging, we used the 10m
Keck\thinspace II telescope with the NIRC2 facility adaptive optics
(AO) imager to obtain natural guide star adaptive optics imaging and
non-redundant aperture mask interferometry (NRM) in the $K^\prime$
filter ($\lambda = 2.124\thinspace \mu$m) on 2016 June 16 UT. These
observations followed the standard observing strategy described by
\cite{Kraus:2016} and previously reported for Ruprecht\thinspace 147
targets by \cite{Torres:2018}. For \epic, we obtained a short sequence
of 6 images and 12 interferograms in vertical angle mode. In both
cases, calibrators were drawn from the other Ruprecht\thinspace 147
members observed on the same night.

The images were analyzed following the methods described by
\cite{Kraus:2016} To briefly recap, the primary star point spread
function (PSF) was subtracted using both an azimuthal median profile
and an independent calibrator that most closely matches the speckle
pattern. Within each image, the residual fluxes as a function of
position were measured in apertures of radius 40\thinspace mas,
centered on each pixel, and the noise was estimated from the RMS of
fluxes within concentric rings around the primary star. Finally, the
detections and detection limits were estimated from the flux-weighted
sum of the detection significances in the stack of all images, and any
location with a total significance greater than 6$\sigma$ was visually
inspected to determine if it was a residual speckle or cosmic ray. No
candidates remained after this visual inspection.

The interferograms were analyzed following the methods described by
\cite{Kraus:2008} and \cite{Ireland:2013}. We Fourier-transformed the
interferograms to extract the complex visibilities, and from those we
computed the corresponding closure phases for each triplet of
baselines. We calibrated the closure phases with other observations of
targets nearby on the sky and in time, and then fit the calibrated
closure phases with binary source models to search for significant
evidence of a companion, but did not find any. We determined the
detection limits using a Monte Carlo process that randomizes the phase
errors and determines the distribution of possible binary fits,
indicating the 99.9\% upper limit on companions in bins of projected
separation.

Within the observations taken with Keck\thinspace II/NIRC2 on 2016
June 16, some targets showed evidence of variable AO correction,
possibly tied to variable seeing over the course of the night. In
particular, some observations of \epic\ displayed elongation of the
PSF core that results when the gain in the AO system does not adapt
quickly enough to changes in seeing. This resulted in little impact on
the imaging limits on wide separations, but did appear to impact the
sensitivity of the NRM observations.

To verify that there were no companions, we therefore made additional
near-infrared high-resolution observations on 2019 June 14 UT with the
PHARO instrument on the Palomar Observatory 5m telescope, behind the
natural guide star AO system. We used a 5-point quincunx dither
pattern that is standard in this type of observation. The dither
pattern step size was 5\arcsec\ and was repeated three times, with
each dither offset from the previous dither by 0\farcs5.

\begin{figure}
\epsscale{1.15}
\plotone{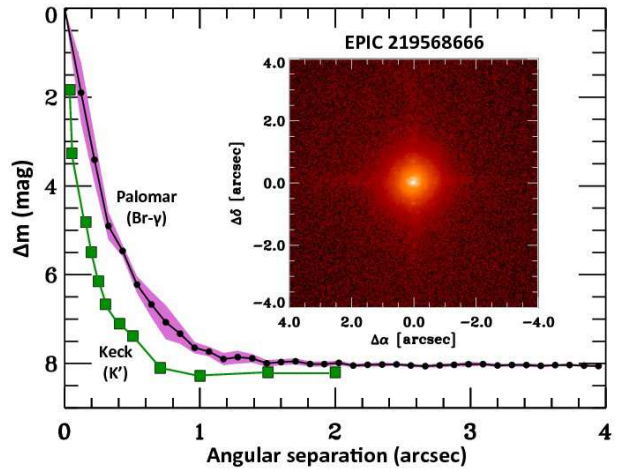}

\figcaption{Companion sensitivity for our adaptive optics imaging of
  \epic.  The black points represent the 5$\sigma$ limits from the
  Palomar Br-$\gamma$ observations and are separated in steps of one
  FWHM ($\sim$0\farcs1); the purple shaded area represents the
  azimuthal dispersion (1$\sigma$) of the contrast determinations (see
  text). The inset image of the target shows no additional companions
  within 4\arcsec. The green squares represent the limits from the
  NIRC2 Keck observations at $K^{\prime}$, which also revealed no
  companions.\label{fig:ao}}

\end{figure}

The images were made in the narrow-band Br-$\gamma$ filter ($\lambda =
2.1686\thinspace \mu$m; $\Delta\lambda = 0.0326\thinspace \mu$m) with
an integration time of 20 seconds and one coadd per frame for a total
of 300 seconds on target.  The camera was in the narrow-angle mode
with a full field of view of $\sim$25\arcsec\ and a pixel scale of
approximately $0\farcs025\thinspace {\rm pixel}^{-1}$. We detected no
additional stellar companions to within a resolution of $\sim$0\farcs1
FWHM (see Figure~\ref{fig:ao}).

The sensitivity of the final combined Palomar AO image was determined
by injecting simulated sources azimuthally around the target every
45\arcdeg\ at separations of integer multiples of the central source's
FWHM \citep{Furlan:2017}. The brightness of each injected source was
scaled until standard aperture photometry detected it with 5$\sigma$
significance. The resulting brightness of the injected sources
relative to the target set the contrast limits at that injection
location. The final 5$\sigma$ limit at each separation was determined
from the average of all of the determined limits at that separation
and the uncertainty on the limit was set by the RMS dispersion of the
azimuthal slices at a given radial distance.

The sensitivity curves from the Keck\thinspace II/NIRC2 and
Palomar/PHARO observations are displayed in Figure~\ref{fig:ao} along
with an inset image zoomed in on the target, showing no other
companion stars.


\subsection{Spectroscopy}
\label{sec:spectroscopyy}

Spectroscopic monitoring of \epic\ was carried out between 2016 May
and 2018 June with the Tillinghast Reflector Echelle Spectrograph
\citep[TRES;][]{Szentgyorgyi:2007, Furesz:2008} on the 1.5m
Tillinghast reflector at the Fred L.\ Whipple Observatory on Mount
Hopkins (Arizona, USA). We gathered a total of 25 spectra at a
resolving power of $R \approx 44,000$ covering the wavelength region
3800--9100~\AA\ in 51 orders. For the order centered at
$\sim$5187~\AA\ containing the \ion{Mg}{1}~b triplet the
signal-to-noise ratios range from 17 to 42 per resolution element of
6.8~\kms.

Radial velocities were measured with TODCOR, a two-dimensional
cross-correlation technique introduced by \cite{Zucker:1994}. A
template matching each component was selected from a pre-computed
library of synthetic spectra based on model atmospheres by
R.\ L.\ Kurucz, and a line list tuned to better match the spectra of
real stars \citep[see][]{Nordstrom:1994, Latham:2002}. These templates
cover a limited wavelength region of $\sim$300~\AA\ centered at
5187~\AA.

For the primary star we estimated the effective temperature ($T_{\rm
  eff}$) and projected rotational velocity ($v \sin i$) by running
grids of one-dimensional cross-correlations of the observed spectra
against synthetic spectra, following \cite{Torres:2002}. The grids
were run over a broad range in those parameters (4500--7500~K,
0--30~\kms) at fixed values of the surface gravity ($\log g$) and
metallicity ([Fe/H]), ignoring the presence of the secondary star
because it is very faint (only about 4\% of the brightness of the
primary; see below). We then chose the combination giving the highest
value of the cross-correlation coefficient averaged over all 25
spectra, weighted by the strength of each exposure. We repeated this
for $\log g$ values of 4.0 and 4.5, bracketing our best estimate
reported later, and [Fe/H] values of 0.0 and +0.5 on either side of
the known cluster abundance. By interpolation we obtained estimates of
$T_{\rm eff} = 6140$~K and $v \sin i = 6~\kms$, with estimated
uncertainties of 100~K and 2~\kms, respectively. These uncertainties
are based on the scatter from the individual spectra, conservatively
increased to account for possible systematic errors.  The temperature
corresponds approximately to a spectral type of F8. For the velocity
determinations we adopted primary template parameters of 6250~K and
6~\kms, the nearest in our grid, along with $\log g = 4.5$ and ${\rm
  [Fe/H]} = 0.0$.

While the presence of the secondary does not affect the above
determinations, its faintness does prevent us from obtaining direct
estimates of its temperature and $v \sin i$ from our spectra. We
therefore adopted a value of $T_{\rm eff} = 4500$~K appropriate for
its mass (typical of a mid-K star), and $v \sin i = 4~\kms$. The
surface gravity and metallicity for the secondary template were taken
to be the same as for the primary.

\begin{figure}
\epsscale{1.15}
\plotone{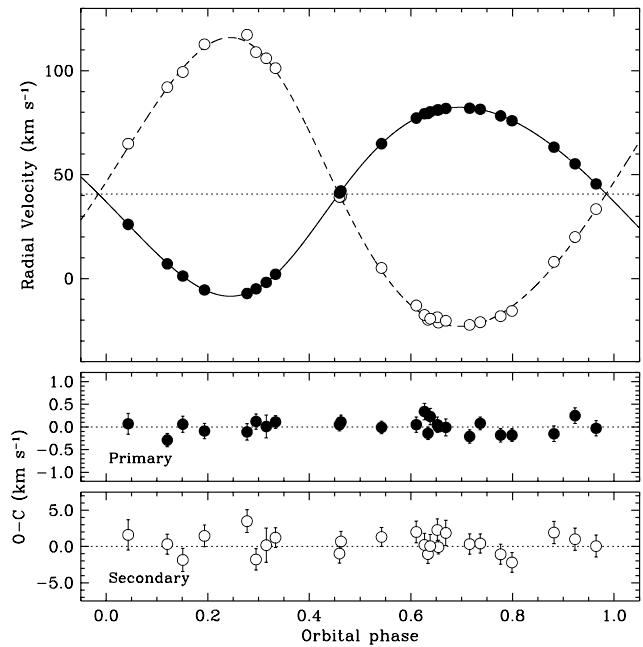}
\figcaption{Radial-velocity measurements for \epic\ with our adopted
  model. Primary and secondary measurements are represented with
  filled and open circles, respectively. The dotted line marks the
  center-of-mass velocity of the system. Error bars are too small to
  be visible. They are seen in the lower panels, which display the
  residuals. Phases are counted from the reference time of primary
  eclipse.\label{fig:rvs}}
\end{figure}

The heliocentric radial velocities measured with TODCOR are presented
in Table~\ref{tab:rvs} along with their uncertainties. We also
determined the average flux ratio at the mean wavelength of our
observations (5187~\AA), which is $\ell_2/\ell_1 = 0.044 \pm 0.003$. A
graphical representation of the velocities is shown in
Figure~\ref{fig:rvs}, along with our final model described below. The
orbit is clearly eccentric. We note that two observations near phase
0.46 (the first and third in Table~\ref{tab:rvs}) are safely outside
of eclipse, and are thus not affected by the Rossiter-McLaughlin
effect.

\setlength{\tabcolsep}{3pt}  
\begin{deluxetable}{lccccc}
\tablewidth{0pc}
\tablecaption{Heliocentric Radial-velocity Measurements of \epic \label{tab:rvs}}
\tablehead{
\colhead{HJD} &
\colhead{RV$_1$} &
\colhead{$\sigma_1$} &
\colhead{RV$_2$} &
\colhead{$\sigma_2$} &
\colhead{Orbital}
\\
\colhead{(2,400,000$+$)} &
\colhead{(\kms)} &
\colhead{(\kms)} &
\colhead{(\kms)} &
\colhead{(\kms)} &
\colhead{Phase}
}
\startdata
 57528.9254  &     \phs42.17  &   0.15  &    \phs\phn39.43  &   1.39  &  0.4623 \\
 57550.9028  &   \phn$-$4.96  &   0.17  &       \phs108.99  &   1.48  &  0.2951 \\
 57552.8726  &     \phs41.13  &   0.14  &    \phs\phn39.26  &   1.31  &  0.4594 \\
 57553.8662  &     \phs64.90  &   0.14  & \phs\phn\phn5.10  &   1.29  &  0.5422 \\
 57554.9623  &     \phs79.50  &   0.14  &       \phn$-$19.80  &   1.28  &  0.6336 \\
 57555.9456  &     \phs81.95  &   0.15  &       \phn$-$22.28  &   1.39  &  0.7157 \\
 57556.9458  &     \phs75.93  &   0.14  &       \phn$-$15.55  &   1.36  &  0.7991 \\
 57557.9371  &     \phs63.21  &   0.17  & \phs\phn\phn8.08  &   1.53  &  0.8817 \\
 57558.9318  &     \phs45.53  &   0.17  &    \phs\phn33.42  &   1.52  &  0.9647 \\
 57566.8695  &     \phs79.32  &   0.18  &       \phn$-$17.51  &   1.58  &  0.6266 \\
 57583.8587  &     \phs26.07  &   0.23  &    \phs\phn64.89  &   2.08  &  0.0434 \\
 57584.7819  &  \phn\phs7.08  &   0.15  &    \phs\phn92.09  &   1.38  &  0.1204 \\
 57700.5724  &     \phs78.30  &   0.15  &       \phn$-$18.06  &   1.39  &  0.7766 \\
 57705.5734  &     \phn$-$5.50  &   0.17  &       \phs112.75  &   1.51  &  0.1937 \\
 57706.5771  &     \phn$-$7.17  &   0.18  &       \phs117.29  &   1.57  &  0.2774 \\
 57710.5737  &     \phs77.23  &   0.17  &       \phn$-$12.98  &   1.50  &  0.6107 \\
 57854.9921  &     \phs81.17  &   0.10  &       \phn$-$21.17  &   0.93  &  0.6542 \\
 57878.9487  &     \phs81.09  &   0.17  &       \phn$-$18.60  &   1.51  &  0.6521 \\
 57908.9128  &  \phs\phn1.21  &   0.18  &    \phs\phn99.40  &   1.58  &  0.1509 \\
 57910.8838  &     \phn$-$1.82  &   0.25  &       \phs106.00  &   2.35  &  0.3153 \\
 57939.9190  &     \phs81.46  &   0.14  &       \phn$-$20.99  &   1.32  &  0.7366 \\
 58034.6680  &     \phs80.25  &   0.18  &       \phn$-$19.28  &   1.62  &  0.6381 \\
 58274.8628  &     \phs81.86  &   0.19  &       \phn$-$20.29  &   1.74  &  0.6688 \\
 58277.9134  &     \phs55.22  &   0.17  &    \phs\phn19.99  &   1.54  &  0.9232 \\
 58294.8240  &  \phs\phn2.07  &   0.14  &       \phs101.25  &   1.35  &  0.3335
\enddata
\tablecomments{Orbital phases are counted from the reference time of
  primary eclipse. Final velocity uncertainties result from scaling
  the values listed for the primary and secondary by the near-unity
  factors $f_1$ and $f_2$, respectively, from our global analysis
  described in Section~\ref{sec:analysis}.}
\end{deluxetable}
\setlength{\tabcolsep}{6pt}  

\section{Analysis}
\label{sec:analysis}

The light curve analysis of \epic\ was carried out using the {\tt eb}
code of \cite{Irwin:2011}, which is based on the Nelson-Davis-Etzel
binary model \citep{Etzel:1981, Popper:1981} implemented in the
popular EBOP program and its descendants. Further details may be found
in our earlier work \citep{Torres:2018}.  The main adjustable
parameters are the orbital period ($P$) and reference epoch of primary
eclipse ($T_0$, which is strictly the time of inferior conjunction in
this code), the central surface brightness ratio in the
\kepler\ bandpass ($J \equiv J_2/J_1$), the sum of the relative radii
normalized by the semimajor axis ($r_1+r_2$) and their ratio ($k
\equiv r_2/r_1$), the cosine of the inclination angle ($\cos i$), and
the eccentricity parameters $e \cos\omega$ and $e \sin\omega$, with
$e$ being the eccentricity and $\omega$ the longitude of
periastron. We adopted a quadratic limb-darkening law for this work,
with coefficients $u_1$ and $u_2$ for each star.

Our detrending procedure is designed to eliminate the obvious
modulation in the light curve due to spots, as well as any long-term
trends (deemed to be of instrumental origin), and at the same time it
also removes other out-of-eclipse variability including tidal
distortions (ellipsoidal variability) and reflection, which for this
long-period and well-detached system are very small in any case.
Therefore, our modeling was run with the calculation of ellipsoidal
variability and reflection disabled, and we restricted the analysis to
data within 0.02 in phase from the center of each eclipse (about 2.5
times the total duration of the eclipses). We accounted for the finite
time of integration of the \ktwo\ long-cadence observations by
oversampling the model light curve and then integrating over the
29.4-minute duration of each cadence prior to the comparison with the
observations \citep[see][]{Gilliland:2010, Kipping:2010}.

Our method of solution used the {\tt emcee\/}\footnote{\url
  http://dan.iel.fm/emcee~.} code of \cite{Foreman-Mackey:2013}, which
is a Python implementation of the affine-invariant Markov chain Monte
Carlo (MCMC) ensemble sampler proposed by \cite{Goodman:2010}. We used
100 walkers with chain lengths of 10,000 each, after discarding the
burn-in.  Uniform (non-informative) or log-uniform priors over
suitable ranges were adopted for all parameters (see below), and
convergence of the chains was checked visually, requiring also a
Gelman-Rubin statistic of 1.05 or smaller for each parameter
\citep{Gelman:1992}.

Flux contamination from neighboring stars was accounted for by
including a third light parameter in our model ($\ell_3$), defined
such that $\ell_1 + \ell_2 + \ell_3 = 1$, in which the values of
$\ell_1$ and $\ell_2$ for this normalization correspond to the light
at first quadrature (phase 0.2235 in this system). An estimate of
$\ell_3$ was obtained from the brightness measurements in
Table~\ref{tab:CFHT} of all companions within the photometric
aperture, using the magnitudes in the CFHT $r$ band, which is close to
the \kepler\ bandpass. To guard against possible errors arising from
the slight bandpass difference, we conservatively inflated the
companion magnitude uncertainties to be no less than 0.2 mag. In this
way we obtained $\ell_3 = 0.026 \pm 0.006$, which we used as a
Gaussian prior in our MCMC analysis. All companions near the edge of
the circular aperture are very faint, so the result is insensitive to
the treatment of partial pixels for those stars.

In eccentric orbits such as that of \epic\ it is usually the case that
$e \cos\omega$ is tightly constrained by the light curve (from the
location of the secondary eclipse), but $e \sin\omega$ is not. The
reverse is true of the radial-velocity curves, making it beneficial to
combine the two types of observations into a joint solution. We took
this approach, solving for three additional adjustable parameters: the
center-of-mass velocity ($\gamma$), and the velocity semiamplitudes of
the primary and secondary ($K_1$ and $K_2$). We handled the relative
weighting between the photometry and the primary and secondary
velocities by including additional adjustable multiplicative
parameters ($f_{\ktwo}$, $f_1$, and $f_2$, respectively) to rescale
the observational errors. These scale factors were solved for
self-consistently and simultaneously with the other orbital quantities
\citep[see][]{Gregory:2005}. The initial error assumed for the
photometric measurements is 200 parts per million (ppm), and the
initial errors for the velocities are those listed in
Table~\ref{tab:rvs}.

For completeness we chose to account for light travel time across the
binary, which can contribute to the displacement of the secondary
eclipse from phase 0.5, although the effect is negligible in this
case: a delay of 26~s, more than three orders of magnitude smaller
than the measured displacement corresponding to about 54,800~s.
Strictly speaking, then, our $T_0$ is referred to the barycenter of
the binary.

As shown below the eclipses of \epic\ are partial and quite shallow
($\sim$11\% and $\sim$4\%). With the expectation that the radius ratio
$k$ would be poorly constrained, as it often is in such cases, we
chose to take advantage of our spectroscopic measurement of the
average light ratio to help constrain $k$ indirectly, given that the
two quantities are strongly correlated ($\ell_2/\ell_1 \propto
k^2$). Our measured value of $\ell_2/\ell_1 = 0.044 \pm 0.003$ at a
mean wavelength of 5187~\AA\ was transformed to the \kepler\ band by
using synthetic spectra based on PHOENIX models taken from the grid of
\cite{Husser:2013}. For this we adopted temperatures of 6100~K and
4500~K, close to those of the components, and a preliminary estimate
of the radius ratio ($k = 0.60$) with which we are able to reproduce
the measured light ratio at 5187~\AA. The result, $\ell_2/\ell_1 =
0.084 \pm 0.005$, was then applied as a prior in our analysis.

Initial solutions revealed that the second-order limb-darkening
coefficients $u_2$ were poorly constrained for both stars. This is to
be expected given the grazing orientation of the system (see
below). Therefore, for the remainder of this work they were held fixed
at the theoretical values tabulated by \cite{Claret:2011},
interpolated according to the adopted temperatures, the metallicity,
and our final surface gravities reported below. The values of $u_2$
adopted for the \kepler\ band are 0.295 and 0.075 for the primary and
secondary, respectively, and are based on ATLAS model atmospheres and
the least-squares procedure favored by those authors for calculating
the coefficients.

\begin{figure}
\centering
\begin{tabular}{c}
\includegraphics[width=8.0cm]{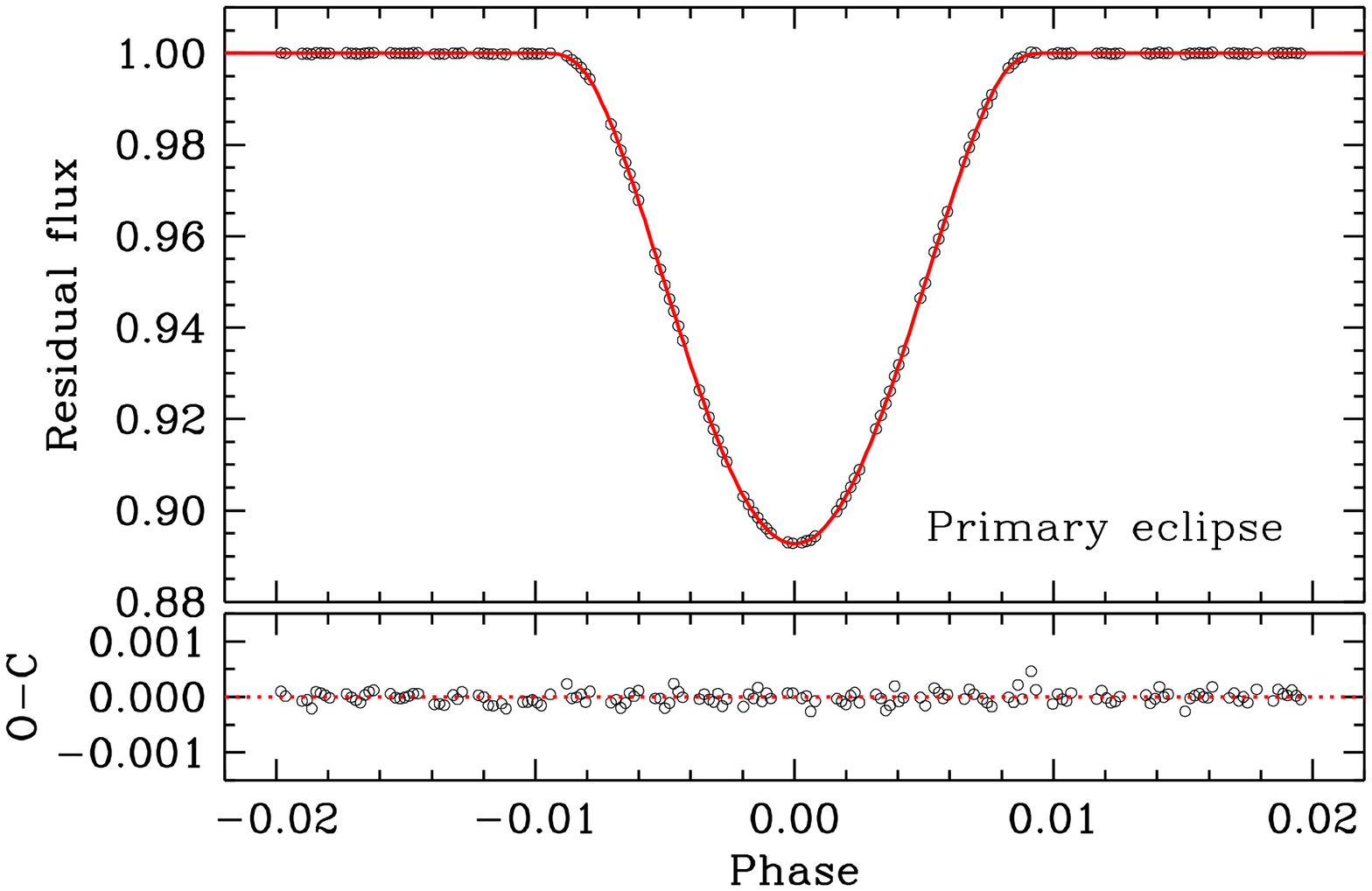} \\ [0.5ex]
\includegraphics[width=8.0cm]{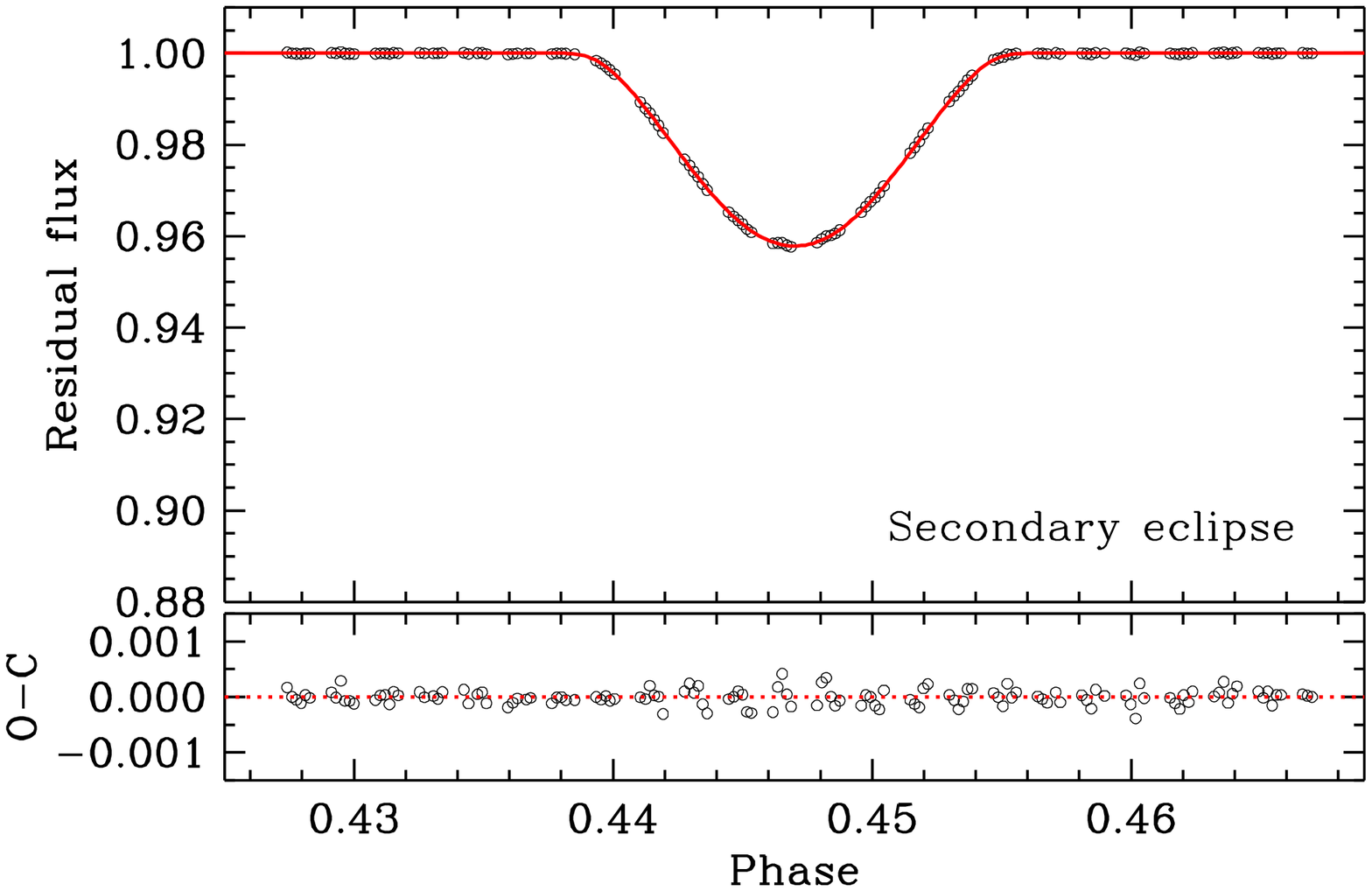} \\ [1ex]
\includegraphics[width=8.0cm]{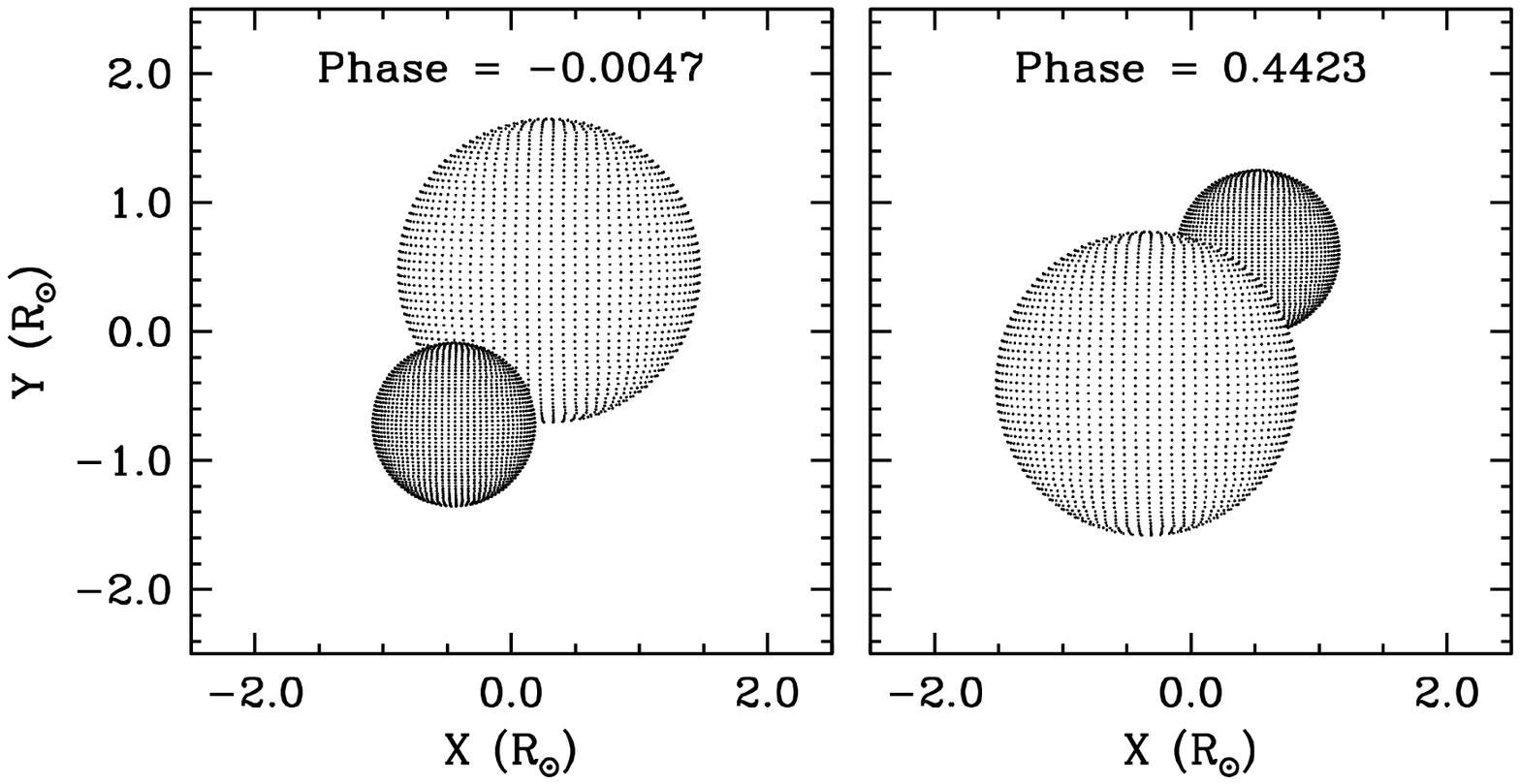}
\end{tabular}
\figcaption{\emph{Top two panels:} \ktwo\ observations of \epic\ and
  our adopted model.  Residuals are shown at the bottom for each
  eclipse. \emph{Bottom:} Illustration of the grazing configuration of
  the system near each of the eclipses, as projected on the
  sky (star sizes drawn to scale). \label{fig:LCfit}}
\end{figure}

We report the results of our analysis in Table~\ref{tab:LCfit}, where
the values given correspond to the mode of the posterior
distributions. Posterior distributions of the derived quantities
listed in the bottom section of the table were constructed directly
from the MCMC chains of the adjustable parameters involved. Our
adopted model along with the \ktwo\ observations can be seen in
Figure~\ref{fig:LCfit}, together with an illustration of the fairly
grazing configuration of the system. The residuals, with an overall
scatter of about 120~ppm, appear slightly larger in the secondary
eclipse.  This may be due to spottedness, which would not be
unexpected for a mid-K star such as the secondary.

\setlength{\tabcolsep}{4pt}
\begin{deluxetable}{lcc}
\tablewidth{0pc}
\tablecaption{Results from our Combined MCMC Analysis for \epic \label{tab:LCfit}}
\tablehead{ \colhead{~~~~~~~~~Parameter~~~~~~~~~} & \colhead{Value} & \colhead{Prior} }
\startdata
 $P$ (days)\dotfill               &  $11.991313^{+0.000013}_{-0.000013}$ & [11, 13] \\ [1ex]
 $T_0$ (HJD$-$2,400,000)\dotfill  &  $57727.23362^{+0.00042}_{-0.00042}$ & [57725, 57729] \\ [1ex]
 $J$\dotfill                      &  $0.2707^{+0.0051}_{-0.0051}$        & [0.02, 1.0] \\ [1ex]
 $r_1+r_2$\dotfill                &  $0.06715^{+0.00045}_{-0.00045}$     & [0.01, 0.20] \\ [1ex]
 $k$\dotfill                      &  $0.544^{+0.016}_{-0.016}$           & [0.1, 1.0] \\ [1ex]
 $\cos i$\dotfill                 &  $0.04128^{+0.00064}_{-0.00064}$     & [0, 1] \\ [1ex]
 $e \cos\omega$\dotfill           &  $-0.083091^{+0.000013}_{-0.000013}$ & [$-$1, 1] \\ [1ex]
 $e \sin\omega$\dotfill           &  $-0.0747^{+0.0010}_{-0.0010}$       & [$-$1, 1] \\ [1ex]
 $\ell_3$\dotfill                    &  $0.0261^{+0.0060}_{-0.0061}$        & [0.0, 0.3]\\ [1ex]
 $\gamma$ (\kms)\dotfill          &  $+40.732^{+0.033}_{-0.033}$         & [30, 50] \\ [1ex]
 $K_1$ (\kms)\dotfill             &  $45.440^{+0.044}_{-0.044}$          & [20, 80] \\ [1ex]
 $K_2$ (\kms)\dotfill             &  $69.46^{+0.37}_{-0.37}$             & [20, 80] \\ [1ex]
 $\ln f_{\ktwo}$\dotfill          &  $-0.479^{+0.047}_{-0.044}$          & [$-5$, 0.9] \\ [1ex]
 $\ln f_1$\dotfill                &  $+0.09^{+0.17}_{-0.14}$             & [$-5$, 5] \\ [1ex]
 $\ln f_2$\dotfill                &  $-0.04^{+0.15}_{-0.14}$             & [$-5$, 5] \\ [1ex]
 Primary $u_1$\dotfill            &  $0.291^{+0.064}_{-0.064}$           & [0, 1] \\ [1ex]
 Secondary $u_1$\dotfill          &  $0.455^{+0.063}_{-0.064}$           & [0, 1] \\ [1ex]
\noalign{\hrule} \\ [-1.5ex]
\multicolumn{3}{c}{Derived quantities} \\ [1ex]
\noalign{\hrule} \\ [-1.5ex]
 $r_1$\dotfill                    &  $0.04349^{+0.00021}_{-0.00022}$     & \nodata \\ [1ex]
 $r_2$\dotfill                    &  $0.02364^{+0.00061}_{-0.00059}$     & \nodata \\ [1ex]
 $i$ (degree)\dotfill             &  $87.634^{+0.037}_{-0.036}$          & \nodata \\ [1ex]
 $\ell_2/\ell_1$\dotfill          &  $0.0781^{+0.0039}_{-0.0037}$        & \nodata \\ [1ex]
 $e$\dotfill                      &  $0.11176^{+0.00069}_{-0.00069}$     & \nodata \\ [1ex]
 $\omega$ (degree)\dotfill        &  $221.98^{+0.38}_{-0.41}$            & \nodata \\ [1ex]
 $f_{\ktwo}$\dotfill              &  $0.620^{+0.029}_{-0.027}$           & \nodata \\ [1ex]
 $f_1$\dotfill                    &  $1.09^{+0.20}_{-0.14}$              & \nodata \\ [1ex]
 $f_2$\dotfill                    &  $0.95^{+0.16}_{-0.12}$              & \nodata \\ [1ex]
 Prim.\ duration (days)\dotfill   &  $0.20800^{+0.00047}_{-0.00047}$     & \nodata \\ [1ex]
 Sec.\ duration (days)\dotfill    &  $0.19582^{+0.00062}_{-0.00062}$     & \nodata \\ [1ex]
 Prim.\ impact param.\dotfill     &  $1.012^{+0.020}_{-0.019}$           & \nodata \\ [1ex]
 Sec.\ impact param.\dotfill      &  $0.872^{+0.017}_{-0.017}$           & \nodata
\enddata
\tablecomments{The values listed correspond to the mode of the
  respective posterior distributions, and the uncertainties represent
  the 68.3\% credible intervals that include a contribution from extra
  photometric noise possibly caused by stellar activity (see
  text). All priors are uniform over the specified ranges, except
  those for $f_{\ktwo}$, $f_1$, and $f_2$, which are
  log-uniform. Eclipse durations are counted from first to last
  contact.}
\end{deluxetable}
\setlength{\tabcolsep}{6pt}

To account for the possibility that the increased secondary residuals
(or the fact that we held the second-order limb-darkening coefficients
fixed) may be causing our parameter uncertainties to be
underestimated, we performed a residual permutation exercise as
follows.  We shifted the residuals from our adopted model by an
arbitrary number of time indices, we then added them back into the
model curve at each time of observation (with wrap-around), and we
carried out a new MCMC analysis on this synthetic data set. We
repeated this 50 times. The residual permutation was done separately
for the in-eclipse and out-of-eclipse regions
\citep[see][]{Hartman:2018}. For each new MCMC analysis we
simultaneously perturbed the theoretical quadratic limb-darkening
coefficients by adding Gaussian noise with a standard deviation of
0.10. The scatter (standard deviation) of the resulting distributions
for each parameter was added in quadrature to the internal errors from
our original MCMC analysis to arrive at the final uncertainties
reported in Table~\ref{tab:LCfit}. The simulation errors are larger
than the internal errors for $P$, $T_0$, and the linear limb-darkening
coefficients.

We note, finally, that our analysis reveals strong correlations among
several of the fitted elements, which is not unexpected for a system
of this configuration. The strongest correlations are between $k$ and
$r_1+r_2$ (correlation coefficient +0.921), between $k$ and $\cos i$
(+0.938), between $r_1+r_2$ and $\cos i$ (+0.992), and between $P$ and
$T_0$ (+0.997).

\section{Absolute dimensions}
\label{sec:dimensions}

The physical properties of \epic\ derived from our MCMC analysis are
collected in Table~\ref{tab:dimensions}. The relative uncertainties
for the masses are $\sim$1\% or smaller. The radii are formally good
to 0.6\% and 2.5\% for the primary and secondary, although we argue
below that systematic errors may be larger.

As described earlier, the primary effective temperature was determined
directly from our spectra, while the secondary value is adopted. These
temperatures along with broadband photometry from the literature allow
us to obtain an estimate of the interstellar reddening along the line
of sight. We proceeded as follows. With the values in
Table~\ref{tab:dimensions} the luminosity-weighted spectroscopic mean
temperature of the system is $\langle T_{\rm eff} \rangle_{\rm sp} =
5877 \pm 84$~K. Color indices from published photometry along with
color/temperature calibrations allow a determination of a mean
photometric temperature, $\langle T_{\rm eff} \rangle_{\rm ph}$, which
depends on the overall reddening, $E(B-V)$. We used photometry in the
Johnson, Tycho-2, 2MASS, and Sloan systems \citep{Zacharias:2013,
  Hog:2000, Skrutskie:2006, Henden:2015} to construct 13
non-independent color indices, and derived a temperature for each one
using the calibrations by \cite{Casagrande:2010} and
\cite{Huang:2015}. Following \cite{Torres:2018}, we adjusted the
temperatures from the \cite{Casagrande:2010} calibrations by $-130$~K
to remove a systematic difference compared to those of
\cite{Huang:2015}. The results were then averaged.  The metallicity
terms in these calibrations were calculated using ${\rm [Fe/H]} =
+0.10 \pm 0.04$ \citep{Curtis:2013}, and the reddening corrections
appropriate for each color index were made as prescribed by
\cite{Cardelli:1989}.  We determined the optimal reddening $E(B-V)$ by
varying it until $\langle T_{\rm eff} \rangle_{\rm ph} = \langle
T_{\rm eff} \rangle_{\rm sp}$. In this way we obtained $E(B-V) = 0.126
\pm 0.023$~mag, corresponding to $A_V = 0.391 \pm 0.071$~mag for a
ratio of total to selective extinction of $R_V = 3.1$.  This is in
good agreement with the value of $E(B-V) = 0.112 \pm 0.029$~mag
derived in our earlier study of the Ruprecht\thinspace 147 eclipsing
system \epicfirst, and also with independent estimates for the cluster
by \cite{Bragaglia:2018} and \cite{Olivares:2019}.

The distance to \epic\ was estimated from the luminosities, bolometric
corrections from \cite{Flower:1996} \citep[see also][]{Torres:2010},
our extinction estimate, and an adopted visual magnitude out of
eclipse of $V = 11.859 \pm 0.050$ \citep{Zacharias:2013}. The result,
$d = 300^{+21}_{-20}$~pc, corresponds to a parallax in good accord
with the entry in the {\it Gaia}/DR2 catalog \citep[][see
  Table~\ref{tab:dimensions}]{Gaia:2018}. Finally, given the
$\sim$3~Gyr age of the cluster, the non-zero eccentricity of the
binary ($e \approx 0.112$) is consistent with the expectation from
tidal theory, which predicts an orbital circularization timescale
\citep[formally 630~Gyr; e.g.,][]{Hilditch:2001} that is longer than
the age of the Universe.

\begin{deluxetable}{lcc}
\tablewidth{0pc}
\tablecaption{Physical Properties of \epic \label{tab:dimensions}}
\tablehead{ \colhead{~~~~~~~~~~Parameter~~~~~~~~~~} & \colhead{Primary} & \colhead{Secondary} }
\startdata
 $M$ ($\mathcal{M}_{\sun}^{\rm N}$)\dotfill &  $1.121^{+0.013}_{-0.013}$  &  $0.7334^{+0.0050}_{-0.0049}$ \\ [1ex]
 $R$ ($\mathcal{R}_{\sun}^{\rm N}$)\dotfill &  $1.1779^{+0.0070}_{-0.0070}$  &  $0.640^{+0.017}_{-0.016}$ \\ [1ex]
 $q \equiv M_2/M_1$\dotfill                 &          \multicolumn{2}{c}{$0.6542^{+0.0036}_{-0.0036}$} \\ [1ex]
 $a$ ($\mathcal{R}_{\sun}^{\rm N}$)\dotfill &          \multicolumn{2}{c}{$27.086^{+0.089}_{-0.089}$} \\ [1ex]
 $\log g$ (dex)\dotfill                     &  $4.3457^{+0.0049}_{-0.0048}$  &  $4.690^{+0.022}_{-0.022}$ \\ [1ex]
 $T_{\rm eff}$ (K)\dotfill                  &  6140~$\pm$~100                &  4500~$\pm$~100 \\ [1ex]
 $L$ ($L_{\sun}$)\dotfill                   &  $1.77^{+0.12}_{-0.11}$     &  $0.152^{+0.016}_{-0.015}$ \\ [1ex]
 $M_{\rm bol}$ (mag)\dotfill                &  $4.109^{+0.072}_{-0.072}$     &  $6.78^{+0.11}_{-0.11}$ \\ [1ex]
 $BC_V$ (mag)\dotfill                       &  $-0.027 \pm 0.100$            &  $-0.624 \pm 0.100$ \\ [1ex]
 $M_V$ (mag)\dotfill                        &  $4.14^{+0.12}_{-0.12}$        &  $7.40^{+0.15}_{-0.15}$ \\ [1ex]
 $v_{\rm sync} \sin i$ (\kms)\tablenotemark{a}\dotfill  &  $4.966^{+0.029}_{-0.029}$  &  $2.699^{+0.070}_{-0.067}$ \\ [1ex]
 $v \sin i$ (\kms)\tablenotemark{b}\dotfill &  $6 \pm 2$                 &  4 (adopted) \\ [1ex]
 $E(B-V)$ (mag)\dotfill                     &          \multicolumn{2}{c}{0.126~$\pm$~0.023} \\ [1ex]
 $A_V$ (mag)\dotfill                        &          \multicolumn{2}{c}{0.391~$\pm$~0.071} \\ [1ex]
 Dist.\ modulus (mag)\dotfill               &          \multicolumn{2}{c}{$7.39^{+0.15}_{-0.15}$} \\ [1ex]
 Distance (pc)\dotfill                      &          \multicolumn{2}{c}{$300^{+21}_{-19}$} \\ [1ex]
 $\pi$ (mas)\dotfill                        &          \multicolumn{2}{c}{$3.33^{+0.23}_{-0.22}$} \\ [1ex]
 $\pi_{Gaia/{\rm DR2}}$ (mas)\tablenotemark{c}\dotfill  &          \multicolumn{2}{c}{$3.277 \pm 0.052$}
\enddata
\tablecomments{The masses, radii, and semimajor axis $a$ are expressed
  in units of the nominal solar mass and radius
  ($\mathcal{M}_{\sun}^{\rm N}$, $\mathcal{R}_{\sun}^{\rm N}$) as
  recommended by 2015 IAU Resolution B3 \citep[see][]{Prsa:2016}, and
  the adopted solar temperature is 5772~K (2015 IAU Resolution
  B2). Bolometric corrections are from the work of \cite{Flower:1996},
  with conservative uncertainties of 0.1~mag, and the bolometric
  magnitude adopted for the Sun appropriate for this $BC_V$ scale is
  $M_{\rm bol}^{\sun} = 4.732$ \citep[see][]{Torres:2010}. See text
  for the source of the reddening. For the apparent visual magnitude
  of \epic\ out of eclipse we used $V = 11.859 \pm 0.050$
  \citep{Zacharias:2013}.}
\tablenotetext{a}{Synchronous projected rotational velocity assuming
  spin-orbit alignment.}
\tablenotetext{b}{Measured projected rotational velocity for the primary.}

\tablenotetext{c}{A global parallax zero-point correction of
  $+0.029$~mas has been added to the parallax \citep{Lindegren:2018a},
  and 0.021~mas added in quadrature to the internal error
  \citep[see][]{Lindegren:2018b}.}

\end{deluxetable}

\subsection{Rotation and activity}
\label{sec:rotation}

The light curve of \epic\ presents obvious variations out of eclipse
with a peak-to-peak amplitude of about 2 milli-magnitudes. We
attribute this to rotational modulation by spots, although it is
unclear which star is responsible. While it may well be the brighter
primary, the secondary is likely to be spotted as well, as suggested
by the somewhat larger photometric residuals during secondary eclipse.
In that case, the dilution effect would imply an intrinsic amplitude
from spots on the secondary of about 2.5\% in the \kepler\ band.

\begin{figure}
\epsscale{1.15}
\plotone{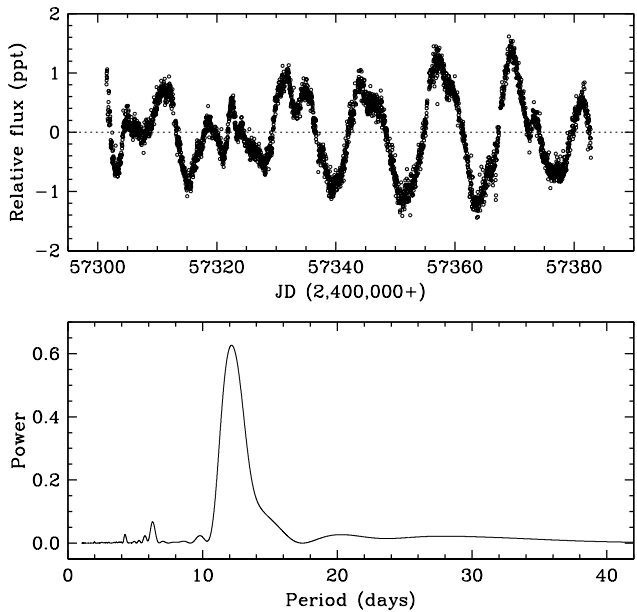}
\figcaption{\emph{Top:} Rotational modulation in the light curve of
  \epic\ (relative flux in parts per thousand, ppt). Eclipses have
  been removed for clarity, along with a long-term instrumental
  drift. \emph{Bottom:} Lomb-Scargle periodogram of the observations
  displaying a dominant peak at $P_{\rm rot} =
  12.2^{+1.1}_{-0.9}$~days. \label{fig:rotation}}
\end{figure}

The observations after removal of the eclipses and a long-term drift
are presented in Figure~\ref{fig:rotation}, along with the
corresponding Lomb-Scargle periodogram.  The peak location at $P_{\rm
  rot} = 12.2^{+1.1}_{-0.9}$~days (uncertainties estimated from the
half width at half maximum) is consistent with the orbital period,
suggesting synchronous rotation.  Our spectroscopically measured
projected rotational velocity of the primary star ($v \sin i = 6 \pm
2~\kms$) agrees with the predicted value of $v_{\rm sync} \sin i$
shown in Table~\ref{tab:dimensions}.  Both of these indications are
consistent with the expected synchronization timescale of
$\sim$300~Myr \citep[e.g.,][]{Hilditch:2001}, given the $\sim$3~Gyr
age of the parent Ruprecht\thinspace 147 cluster.

Aside from the small photometric modulation described above, the
activity level of the binary appears to be low, as we see no
indicators of such activity in our spectra, nor does the object appear
to have been detected in X rays.

\section{Comparison with theory}
\label{sec:models}

\begin{figure}
\epsscale{1.15}
\plotone{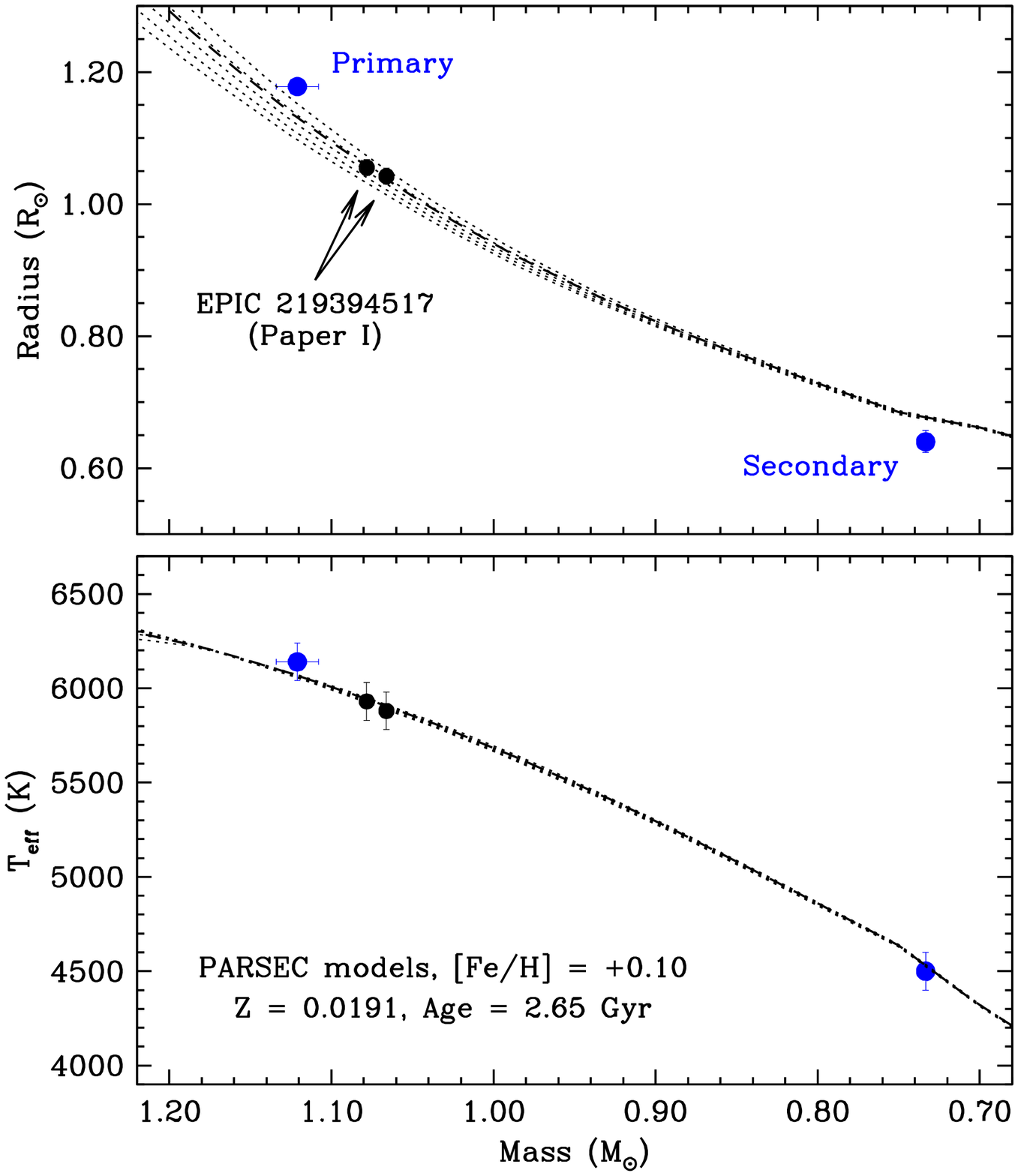}

\figcaption{Comparison of the measured masses, radii, and effective
  temperatures of \epic\ (blue points) against stellar evolution
  models from the PARSEC series \citep{Chen:2014}. We show also the
  results for the \epicfirst\ system (smaller black points) from our
  earlier study \citep{Torres:2018}. The adopted metallicity is that
  of the Ruprecht\thinspace 147 cluster, ${\rm [Fe/H]} =
  +0.10$. Dotted lines correspond to model isochrones from 2 to 3~Gyr
  in equally spaced logarithmic intervals, and the heavier dashed line
  represents the best-fit age of 2.65~Gyr from our earlier work on
  \epicfirst. The error bars shown are the nominal values from
  Table~\ref{tab:LCfit}, which do not reflect correlations discussed
  in the text. \label{fig:MR}}
\end{figure}

Our precise determinations of the masses and radii of \epic, and the
increased leverage afforded by a mass ratio significantly different
from unity, offer an opportunity for a valuable comparison with
current stellar evolution models. They also permit an independent
estimate of the age of the Ruprecht\thinspace 147 cluster to
supplement the one from our previous study of
\epicfirst\ \citep{Torres:2018}.  Figure~\ref{fig:MR} displays the
measured radius and temperature of the components as a function of
their measured mass, together with model isochrones from the PARSEC
series \citep{Chen:2014} computed for the cluster metallicity of ${\rm
  [Fe/H]} = +0.10$. For reference we include the results for
\epicfirst\ from Paper~I. The model isochrones are the same as shown
in Figure~7 of that work, and span ages between 2 and 3~Gyr. The heavy
dashed line represents the 2.65~Gyr isochrone that provided the best
fit to \epicfirst\ in our earlier study.

While the isochrones match the effective temperature of the primary of
\epic\ to within its uncertainty (the secondary $T_{\rm eff}$ is
adopted from models), the radius of the primary appears $\sim$4\%
larger than expected for its mass (6.7$\sigma$), if we take the
best-fitting model from \epicfirst\ as the reference. Furthermore, the
radius of the secondary is about 6\% smaller than the models predict
(2.2$\sigma$), assuming the mass is accurate.  These differences are
somewhat surprising, particularly for the primary, given the high
precision of the observations and the relatively straightforward
analysis, and the expectation of fewer complications from the longer
orbital period compared to our previous study of the more active
6.5-day system \epicfirst. Repeating the comparison with the MIST
models \citep{Choi:2016} gives essentially the same result.

A casual look at Figure~\ref{fig:MR} may give the impression that
relatively small shifts in the locations of the primary and secondary
in just the right directions would bring satisfactory agreement with
the reference model from \epicfirst. For instance, an increase in
$M_1$ by about 2$\sigma$ together with an increase in $R_2$ also by
about 2$\sigma$ would be sufficient to obtain a reasonably good fit.
However, this ignores the strong correlations that exist among the
inferred masses and radii, which restrict the shifts one may apply in
each direction if they are to remain within the multi-dimensional
confidence region mapped out by our MCMC analysis. As an example,
$M_1$ and $M_2$ have a correlation coefficient of +0.981, indicating
they cannot be varied independently.

\begin{figure}
\epsscale{1.2}
\plotone{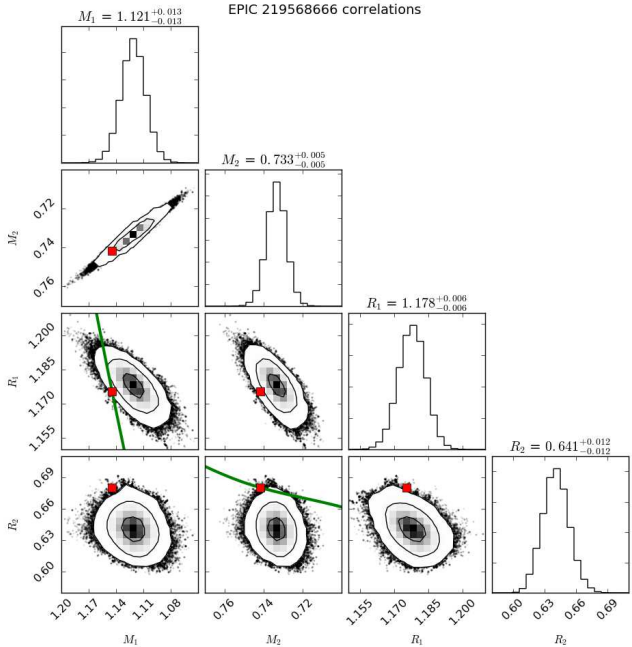}

\figcaption{``Corner'' plot \citep{Foreman-Mackey:2016}\footnote{\url
    https://github.com/dfm/corner.py~.} based on the results of our
  MCMC analysis showing the correlations among the derived masses and
  radii of \epic. The contours correspond to the 1-, 2-, and
  3-$\sigma$ confidence levels. The green diagonal lines in the
  [$M_1$, $R_1$] and [$M_2$, $R_2$] panels are segments of the reference
  isochrone from Figure~\ref{fig:MR}. The red squares mark the points
  on the isochrone with the smallest normalized difference in
  four-dimensional space relative to the measured mass and radius
  values in Table~\ref{tab:LCfit} (see text).\label{fig:corner}}
\end{figure}

A more quantitative assessment of how well the measurements agree with
the reference isochrone that implicitly accounts for all correlations
may be obtained using the chains from our {\tt emcee} analysis, with a
combined length of $10^6$ links. For each link we determined the
normalized distance in four-dimensional parameter space between the
values of [$M_1$, $M_2$, $R_1$, $R_2$] and the nearest pair of points
on the isochrone representing the location of the primary and
secondary.
%
%
We normalized the separation along each axis by the standard deviation
of the corresponding variable as determined from the chains. The
masses and radii on the isochrone that are closest to the link giving
the smallest distance are represented with square symbols in
Figure~\ref{fig:corner}. This figure displays a ``corner'' plot
showing the correlations among the four quantities. The contours
correspond to the 1-$\sigma$, 2-$\sigma$, and 3-$\sigma$ confidence
levels, and the panels for [$M_1$, $R_1$] and [$M_2$, $R_2$] also show
the reference isochrone from Figure~\ref{fig:MR}. The square symbols
in the latter two panels deviate from the values reported in
Table~\ref{tab:LCfit} by more than 2$\sigma$ for the primary and more
than 3$\sigma$ for the secondary. Similar offsets are seen in the
other panels. The overall discrepancy with the model is at about the
3-$\sigma$ level, which we consider significant.

%

If we assume for a moment that the masses of both stars are unbiased,
the positive deviation in radius for the primary star may be
speculated to be due to stellar activity, which is believed to be the
underlying cause of the phenomenon of ``radius inflation'' in
convective stars \citep[see, e.g.,][]{Mullan:2001, Chabrier:2007,
  Torres:2013, Feiden:2013, Feiden:2014, Somers:2015}. We note,
though, that the activity level of the primary seems relatively low,
based on the photometric amplitude of the rotational modulation
(provided it comes from this star and not the secondary) and on the
lack of spectroscopic activity indicators (H$\alpha$ or \ion{Ca}{2} H
and K emission) or X-ray emission. Furthermore, radius inflation is
most often seen in K and M dwarfs, and much more rarely among
higher-mass convective stars near a solar mass, though a few examples
do exist \citep[e.g., the secondary components of CV\thinspace Boo,
  FL\thinspace Lyr, V1061\thinspace Cyg, and V636\thinspace Cen, with
  masses of 0.97, 0.96, 0.93 and 0.85~$M_{\sun}$, respectively;
  see][]{Torres:2008, Helminiak:2019, Torres:2006, Clausen:2009}. None
are as massive as the F8 primary of \epic, however, and all rotate
more rapidly.

On the other hand, we can think of no plausible physical explanation
for the small radius of the secondary at its nominal mass, although
the deviation is admittedly less significant than for the primary.
Instead, we postulate that the discrepancies for both stars are more
likely to be due to systematic errors stemming from our analysis
and/or from the observations themselves. In the next section we
discuss possible sources of these errors, 
and the tests we carried out to
investigate them.

We note here that light curve solutions for systems with shallow
eclipses such as \epic\ can often lead to a biased measure of the
radius ratio (especially, but not necessarily, when the components are
similar, which is not the case here) because of degeneracies among
several orbital elements \citep[see, e.g.,][]{Andersen:1991,
  Torresetal:2010, Kraus:2015}. Such degeneracies are present in our
own analysis, as pointed out at the end of
Section~\ref{sec:analysis}. On the other hand, the sum of the radii is
typically more robust \citep[e.g.,][]{Andersen:1983}. Interestingly,
we estimate that an increase from the value of $k \approx 0.54$ we
determine to about 0.60 would yield very good agreement with the
models shown in Figure~\ref{fig:MR}. This may be taken as
circumstantial evidence of a possible bias in the radius ratio,
although the situation is likely more complex given the correlations
between $k$ and other elements.

\begin{figure}
\epsscale{1.15}
\plotone{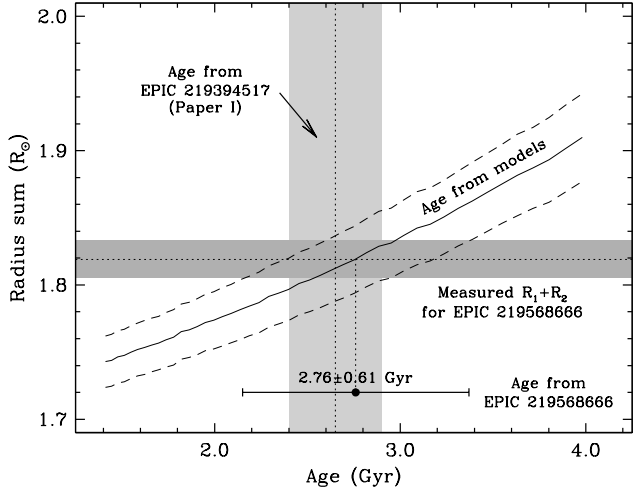}
\figcaption{Age determination for \epic\ based on the radius sum (dot
  and error bars, $2.76 \pm 0.61$~Gyr). The horizontal shaded area
  represents the measured value of $R_1+R_2$ and its uncertainty. The
  age from the PARSEC models along with the 1$\sigma$ uncertainty
  resulting from the errors in the measured masses is shown with the
  solid and dashed lines. The age determined from our earlier study of
  \epicfirst\ \citep{Torres:2018} is shown for reference by the
  vertical shaded area. \label{fig:age}}
\end{figure}

If we set aside for now the individual radii and rely only on the
radius sum ($R_1 + R_2 = 1.819 \pm 0.014~R_{\sun}$), which we expect
to be more accurate, it is still possible to estimate an age for the
system using the same PARSEC models as above. This is illustrated in
Figure~\ref{fig:age}, in which the solid curve gives the age predicted
from theory (for the adopted cluster metallicity of ${\rm [Fe/H]} =
+0.10$) as a function of the radius sum computed at the measured
masses for \epic. The corresponding 1$\sigma$ error interval that
comes from the uncertainty in the measured masses is shown with dashed
lines. The intersection of this curve with the measured radius sum for
\epic\ (horizontal shaded area) leads to an age estimate of $2.76 \pm
0.61$~Gyr (dot and error bars at the bottom). Although the uncertainty
is more than twice that of the estimate for \epicfirst\ from Paper~I
($2.65 \pm 0.25$~Gyr at fixed metallicity), the determinations are
consistent within the smallest of their errors. In both cases we used
the same models. Together the two binaries therefore support an age
for the cluster slightly under 3~Gyr.

\section{Discussion}
\label{sec:discussion}

In this section we examine a number of possible explanations for the
unexpected disagreement illustrated in Figure~\ref{fig:MR} between
models and the mass and radius measurements for \epic, which appears
statistically significant, as discussed above. Barring an unusual
instance of (activity-related?) radius inflation for a star as massive
as the primary, or severely underestimated mass and/or radius
uncertainties that we think is unlikely, we proceed on the assumption
that there may be a subtle bias somewhere in our analysis that is
causing the primary to appear to be too large and the secondary too
small compared to predictions, or perhaps the primary mass to be too
small.  We address potential errors in the spectroscopy, errors in the
priors for the light ratio and third light that we applied in our MCMC
analysis, the possibility of biases in our detrending procedures for
the \ktwo\ photometry, and the impact of the cadence of the
photometric observations.

\subsection{Mass errors}
Earlier we proposed that an underestimate of the radius ratio is one
plausible explanation for the measured slope in the mass-radius
diagram being too steep (Figure~\ref{fig:MR}). In principle a bias in
the mass ratio may contribute as well. This could result from a
mismatch between the templates used in our TODCOR analysis (involving
mainly the $T_{\rm eff}$ parameter, which affects the velocities the
most) and the spectra of the real components. While the primary
temperature is well determined directly from our spectra, the
secondary temperature (4500~K) was adopted from models because that
star is too faint for an independent determination. We repeated the
radial-velocity measurements for a range of temperatures for the
secondary template between 4000 and 5000~K in steps of 250~K, but
found that the residuals from a spectroscopic orbital solution were
worse, the masses changed by less than 1\% compared to the adopted
values, and the mass ratio changed by even less. We conclude from this
that the masses appear to be robust, and are therefore not likely to
be the main source of the discrepancy, though they may still have some
influence. This then shifts the focus to the radii, which depend more
critically on the photometry.

\subsection{Alternate solutions}
A separate MCMC analysis without the radial velocities gave
essentially the same results for both the radius sum and radius ratio,
again suggesting that the problem may lie in the photometry.
Splitting the photometric data set in two produced similar parameters
in each half, from which we conclude any bias is not time-dependent.

\subsection{Priors}
Our adopted solution in Table~\ref{tab:LCfit} used Gaussian priors for
both the light ratio and third light, each based on empirical
constraints. To investigate the effect of these priors, we repeated
the analysis without them as well as using one prior but not the
other.  All three solutions resulted in values of $k$ and $r_1+r_2$
(and also of $\ell_2/\ell_1$ and $\ell_3$) that were very close to those
obtained when applying both priors together, suggesting little or no
tension between the photometry and the external constraints.
Nevertheless, we examined each of the priors more closely.

\subsubsection{Spectroscopic light ratio}
The prior on the light ratio can potentially have the strongest impact
on the inferred value of the radius ratio because of the direct
correlation between them ($\ell_2/\ell_1 \propto k^2$). For example,
grids of MCMC solutions imposing Gaussian light ratio priors ranging
from 0.030 to 0.150 (straddling our adopted value of $0.084 \pm
0.005$) and the same prior on $\ell_3$ as in our original analysis show
indeed that the derived $k$ values change significantly from 0.50 to
0.73 (a 46\% change), while the values of $r_1+r_2$ change in tandem
but less, from about 0.066 to 0.071 (8\%), supporting the notion that
the radius sum is more robust. Removing the $\ell_3$ prior in this
exercise causes the radius sum to change in the opposite direction,
from 0.071 to 0.066. However, in both cases the quality of the
solution as measured by the $f_{\ktwo}$ scale factor for the
photometric errors degrades considerably toward the upper end of the
$\ell_2/\ell_1$ range we explored.

Our adopted light ratio prior in the \kepler\ band is extrapolated
from our measured value at $\sim$5187~\AA, as described in
Section~\ref{sec:analysis}, and as such it is subject to error. The
extrapolation used synthetic spectra by \cite{Husser:2013} appropriate
for each component in order to calculate the wavelength dependence of
the flux ratio. The normalization was performed using $k = 0.60$,
which is the value of the radius ratio we find is needed in order to
reproduce the measured $\ell_2/\ell_1$ value at 5187~\AA\ ($0.044 \pm
0.003$).  Interestingly, this value of $k$ (which is independent of
our MCMC analysis) is also the one that seems to provide a match
between the PARSEC evolutionary models and the individual masses and
radii of \epic, assuming $r_1+r_2$ is accurate. As noted above,
however, our MCMC analysis returns a lower $k$ value close to 0.54,
not 0.60.

As a check on our extrapolation we repeated the light ratio
determination with TODCOR in other echelle orders available in our
spectra sampling the entire \kepler\ bandpass. Because our standard
template library \citep{Nordstrom:1994, Latham:2002} only spans about
300~\AA\ around the region of the \ion{Mg}{1}~b triplet, we used
synthetic spectra based on PHOENIX models from the same
\cite{Husser:2013} library as above that cover the entire optical
range. We note that these synthetic spectra were not used for our
original radial-velocity determinations because at high resolution
they do not match real stars as well as the templates from our own
library, which are based on a line list tuned for that
purpose. Nevertheless, the \cite{Husser:2013} library suffices for a
determination of the light ratio, which does not require high
resolution.

\begin{figure}
\epsscale{1.15}
\plotone{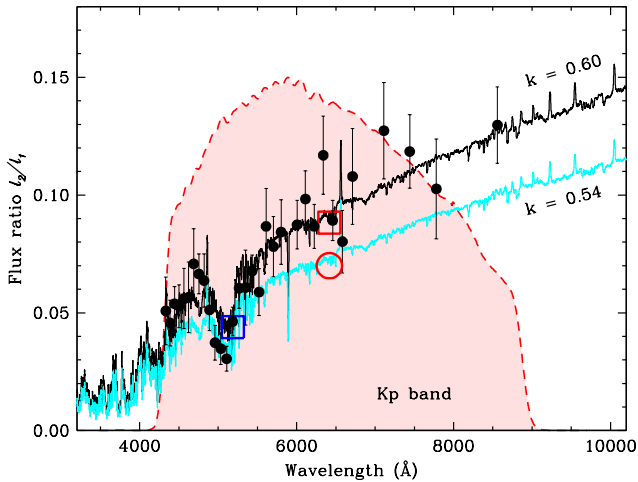}

\figcaption{Spectroscopic flux ratio measurements (points with error
  bars) compared with calculations based on synthetic spectra for
  solar-metallicity from \cite{Husser:2013}, for component
  temperatures of 6100~K and 4500~K, and $\log g = 4.5$. Two model
  predictions are shown. One (black) was normalized using a radius
  ratio of $k = 0.60$, a value such that the curve integrated over the
  spectral order containing the \ion{Mg}{1}~b triplet returns a value
  (blue square) matching the measured light ratio at 5187~\AA. The
  other model (cyan) was normalized using a radius ratio of $k =
  0.54$, which is the value resulting from our MCMC analysis
  (Table~\ref{tab:LCfit}).  The predictions of both models for the
  \kepler\ band are represented by the red open square and circle,
  respectively. The \kepler\ response function is also
  shown. \label{fig:fluxratio}}
\end{figure}

We measured $\ell_2/\ell_1$ in 31 spectral orders between
4300~\AA\ and 8900~\AA\ that have high enough flux and are
sufficiently free from telluric lines. The results are displayed in
Figure~\ref{fig:fluxratio} as points with error bars. Also shown is
the predicted flux ratio from the synthetic spectra for two different
values of the radius ratio. The top curve is for $k = 0.60$, the value
required to reproduce the measured light ratio at 5187~\AA, and the
bottom curve is for $k = 0.54$, which is our result from the light
curve analysis. The open squares on the top, $k = 0.60$ curve mark the
predicted values at 5187~\AA\ (matching the spectroscopic $0.044 \pm
0.003$ measurement by construction) and at the \kepler\ band ($0.084
\pm 0.005$). They were computed by integrating the theoretical flux
ratio over the corresponding response functions (assumed to have a
top-hat shape for the \ion{Mg}{1}~b order).  The \kepler\ band
prediction for the $k = 0.54$ curve is marked with an open circle. For
reference the shaded area represents the response function for
\kepler, with arbitrary normalization.

Our order-by-order light ratio measurements over the \kepler\ band
clearly agree best with the upper ($k = 0.60$) curve in
Figure~\ref{fig:fluxratio}, and thus support the accuracy of the
$\ell_2/\ell_1$ prior used in our analysis, which was based on that
value of the radius ratio. At the same time, this purely spectroscopic
approach conflicts with the result for $k$ returned by our MCMC
analysis, suggesting the latter value is underestimated. We note that
this conclusion is independent of the stellar evolution models.

\subsubsection{Third light}
Our Gaussian prior for $\ell_3$ ($0.026 \pm 0.006$) relies on brightness
measurements for all known stars within the photometric aperture,
based on a seeing-limited image from the CFHT (Figure~\ref{fig:CFHT}
and Section~\ref{sec:analysis}). While our higher-resolution AO
imaging in Section~\ref{sec:imaging} revealed no additional companions
outside of 0\farcs1, it is still possible we are missing flux due to
even closer companions. To explore this possibility and its effect on
the analysis, we ran a grid of MCMC solutions in which we varied the
$\ell_3$ prior between 0.03 and 0.12, maintaining our light ratio prior
as in our original analysis. We found only a very small increase in
$k$ (0.539--0.544) coupled with a reduction in $r_1+r_2$
(0.0670--0.0654), which together result in an even smaller value for
the secondary radius than indicated in Table~\ref{tab:dimensions}.
Furthermore, if the missing third light were due to a physically
associated single star, a value as high as $\ell_3 = 0.12$ would imply a
brightness for that star relative to the primary of $\ell_3/\ell_1
\approx 0.13$ in the \kepler\ band, which is even brighter than the
secondary star ($\ell_2/\ell_1 = 0.084$). There is no sign of such a
tertiary in our spectra, nor do we see any trend in the
radial-velocity residuals from our spectroscopic orbit that might
indicate an outer companion bound to the binary.

\subsection{Limb darkening}
For the present analysis we adopted the quadratic limb-darkening law
for both stars, which is commonly used when dealing with space-based
photometry. We allowed the first-order coefficients to vary and held
the second-order coefficients fixed from theory because they are
poorly constrained by the data given that the eclipses are grazing.
The first-order coefficient we obtain for the primary, $0.291 \pm
0.064$, is quite consistent with the theoretical value of 0.343 from
\cite{Claret:2011}, while the one for the secondary star, $0.455 \pm
0.064$, is lower than expected (0.674). In principle the use of a
different or possibly more sophisticated limb-darkening prescription
could affect the resulting geometric parameters to some extent,
particularly the radius sum and radius ratio. Unfortunately we are not
able to test alternate formulations here because the public {\tt eb}
code we use restricts our choices to either the linear or quadratic
laws. Nevertheless, as an experiment we repeated the analysis with the
linear law, and found that neither $k$ nor $r_1+r_2$ changed
significantly. The linear coefficients we derived in this case for the
primary and secondary, $0.464 \pm 0.012$ and $0.428 \pm 0.034$, may be
compared with the theoretical values of 0.564 and 0.730,
respectively. Using the linear law for one star and the quadratic law
for the other also did not affect the geometric parameters in any
meaningful way. While these experiments do not completely rule out the
possibility that a different or higher-order limb-darkening formula
might have a more significant impact, the fact that even reverting to
the simpler linear law did not change the results makes that seem less
likely. In retrospect this is not all that surprising, as the grazing
configuration means that the photometry typically has very little
discriminating power on the details of limb darkening.

\subsection{Detrending}
Given the above evidence that neither our spectroscopy nor our
Gaussian priors on $\ell_2/\ell_1$ or $\ell_3$ seem to be causing a bias
in our lightcurve analysis, suspicion falls on the photometry itself,
and in particular on the processing (detrending) to which it was
subjected prior to analysis.

%
%
%
%
%
%

We therefore tested different ways of treating the photometry,
focusing both on the systematics-removal stage and on the removal of
low-frequency variability, to see if residual systematic errors or
subtle biases from our low-frequency removal might explain the
discrepancy between our best-fit radius ratio and the radii predicted
by evolutionary models.

First, we tried removing low-frequency variability from the
\ktwo\ light curve in a different way. Instead of dividing away the
best-fit spline from the light curve solution, we fit for a new spline
after the removal of systematics while manually masking the eclipses.
We then tweaked our systematics removal process to see if subtle
biases from our simultaneous fit to the eclipse shapes, the
\ktwo\ systematics, and low-frequency variability were causing a bias
in the radius ratio.  We tested the light curve with only a first-pass
systematics correction, and found that the best-fit radius ratio from
this light curve was essentially unchanged compared to our original
processing.

We also tested different light curves produced by simultaneously
fitting the systematics and low-frequency variability along with more
sophisticated models for the eclipses. The \cite{Mandel:2002} code we
used in our detrending was designed for modeling the light curves of
transiting planets, and assumes that each set of transits or eclipses
has the same limb-darkening profile, and furthermore, it does not
include a third-light contribution. In our tests we relaxed both of
these assumptions by including a third-light contribution in the
eclipse model and by allowing the flexibility for the code to fit each
set of eclipses with their own quadratic limb darkening
coefficients. Interestingly, none of these changes significantly
affected the best-fit radius ratio $k$ or the radius sum $r_1+r_2$
derived from the light curve.





\subsection{Impact of the radius ratio on the light curve}

In order to quantify the impact that a change in the radius ratio has
on the light curve itself, we compared our adopted model with another
in which we forced $k$ to be approximately 0.60. We did this by
applying a tight prior on $k$ and removing the prior on
$\ell_2/\ell_1$ so as to avoid conflict. All other parameters were
solved for in the same way as before. The difference between the
synthetic light curves from these two models has a maximum amplitude
of about 100~ppm during the ingress/egress of the secondary eclipse,
and less than half of that during the primary eclipse. We note that
this is of the same order as the overall scatter of the
\ktwo\ photometry from our adopted model ($\sim$120~ppm).  We
speculate that this difference may be too small for our analysis to
detect given the photometric noise.

\subsection{Cadence of the photometric observations}

Finally, we explored the possibility that the relatively long 29.4~min
sampling of the \ktwo\ light curve may be negatively impacting our
results, given that only a handful of observations were recorded near
the points of first and last contact (see Figure~\ref{fig:LCfit}),
which are critical for defining the radii. For this we used the same
$k = 0.60$ solution mentioned in the previous section, and oversampled
the corresponding model to emulate the 1~min short cadence (SC)
observations that \ktwo\ has obtained for many targets (but not for
\epic). As a check, we also sampled it at the same 29.4~min rate (long
cadence, LC) as the real observations. We then added Gaussian noise
appropriate to each cadence, and subjected the two synthetic data sets
to identical MCMC analyses as with the real data, integrating the
model over 29.4~min for the LC time series but not for the SC time
series. We repeated this several times, each with different noise, and
found that within the uncertainties both synthetic data sets gave the
same results for $k$ and $r_1+r_2$, on average, and that those results
were also similar to the input values of those parameters for this
experiment. We conclude that the cadence of the photometric
observations has little influence in this particular case.


\subsection{Summary}

From the tests described above we conclude that the apparent bias in
the radius ratio suggested by Figure~\ref{fig:MR} and supported by
Figure~\ref{fig:fluxratio} does not seem to be caused by errors in the
radial velocity measurements, by incorrect priors on $\ell_2/\ell_1$
or $\ell_3$, by the subtleties of our detrending procedures, or by the
finite cadence of the observations.  The change in $k$ required to
match the slope in the mass-radius diagram (from $\sim$0.54 to
$\sim$0.60) is about 11\% of the value we measure. While this change
may seem significant at face value (about 3.7 times the formal
uncertainty in $k$), the impact on the shape of the light curve is
actually very small, as indicated earlier, and it may be that the data
available are insufficient to discern this difference. Another
possibility that cannot be ruled out has to do with the presence of
spots on one or both stars, especially given the grazing configuration
of the \epic\ system. Spots can cause systematic errors in the
measured eclipse depths (changing the surface brightness ratio), which
in turn can lead to biases in other geometric parameters such as the
radius ratio, the inclination angle, and even the radius sum. An
example with a much more detailed discussion of spot effects on
eclipsing binary parameters may be found in the work of
\cite{Irwin:2011}.

\section{Concluding remarks}
\label{sec:remarks}

\epic\ is the second eclipsing binary we have analyzed in the
Ruprecht\thinspace 147 cluster after \epicfirst\ \citep{Torres:2018},
based on \ktwo\ photometry and follow-up spectroscopic
observations. While the formal precision of our mass and radius
determinations is quite high ($\sim$1\% relative errors in the masses,
and about 0.6\% and 2.5\% errors in the radii), the agreement with
stellar evolution models is less satisfactory for \epic\ than it was
for the system we studied earlier. The primary star seems too large
for its mass (a 6.7$\sigma$ discrepancy) while the secondary is a bit
too small (though only by 2.2$\sigma$). Activity-related radius
inflation seems unlikely for the primary for the reasons indicted
earlier, and there is no evidence of any strong systematic errors in
the masses, which leads us to suspect a bias in the radius ratio we
inferred from the \ktwo\ light curve.

A weakly constrained radius ratio is not an uncommon occurrence in
light curve analyses, though it often goes unnoticed. The usual cure
for this problem is the use of an external constraint on the light
ratio derived, e.g., from spectroscopy, as we have done here. For
\epic\ this does not seem to have completely eliminated biases,
despite our best attempts. Careful examination of the critical
ingredients for the mass and radius determinations, as detailed in the
preceding section, has given no clues as to what could be causing $k$
to be underestimated in our analysis. A remaining possibility is a
bias in the geometric parameters caused by spots on one or both
stars. Simulations by \cite{Morales:2010} have shown that these
effects appear to be maximized when the spots are concentrated at the
poles. Although we have no information on the latitudinal distribution
of spots in this system, it is possible that acquiring multi-color
observations might help, as the different sensitivity to spots as a
function of wavelength could help break degeneracies with other
effects such as limb darkening, particularly in a grazing
configuration such as that of the present system.


The example of \epic\ serves as a cautionary tale about the need to be
aware of the potential for biases in the solution of light curves,
especially regarding properties that often poorly constrained such as
the radius ratio. This can even affect photometric measurements from
space-based missions, which may well be internally highly precise but
could still suffer from subtle systematic errors not reflected in the
formal uncertainties. Such errors could result, e.g., from the complex
processing space photometry is typically subjected to, prior to use,
or from physical effects such as the presence of spots, as mentioned
above.  Independent high-precision photometric observations of
\epic\ such as may be obtained in the future with NASA's Transiting
Exoplanet Survey Satellite (TESS), or at other wavelengths from the
ground, may shed some light on this issue.

We have shown that it is still possible to infer a useful estimate of
the age of \epic\ by relying only on the sum of the radii, bypassing
the use of the individual radii involving $k$. The result, $2.76 \pm
0.61$~Gyr at a metallicity fixed to that of the cluster, is consistent
with the independent estimate obtained in our earlier study of
\epicfirst\ ($2.65 \pm 0.25$~Gyr), if much less precise. Together
these two age estimates therefore point to an age for
Ruprecht\thinspace 147 near 2.7~Gyr. We expect to strengthen this
determination even further as studies are completed for three
additional eclipsing binaries in the cluster that are underway.




\begin{acknowledgements}

The spectroscopic observations of \epic\ were gathered with the help
of P.\ Berlind, M.\ Calkins, G.\ Esquerdo, and D.\ Latham. J.\ Mink is
thanked for maintaining the CfA echelle database. We are also grateful
to J.\ Irwin for implementing changes in the {\tt eb} program that
facilitated the present analysis, and to the anonymous referee for
helpful comments and suggestions.  G.T.\ acknowledges partial support
from NASA's Astrophysics Data Analysis Program through grant
80NSSC18K0413, and to the National Science Foundation (NSF) through
grant AST-1509375. J.L.C.\ is supported by the NSF Astronomy and
Astrophysics Postdoctoral Fellowship under award AST-1602662, and by
NASA under grant NNX16AE64G issued through the \ktwo\ Guest Observer
Program (GO~7035).  This work was performed in part under contract
with the California Institute of Technology (Caltech)/Jet Propulsion
Laboratory (JPL) funded by NASA through the Sagan Fellowship Program
executed by the NASA Exoplanet Science Institute.  The research has
made use of the SIMBAD and VizieR databases, operated at the CDS,
Strasbourg, France, and of NASA's Astrophysics Data System Abstract
Service. The research was made possible through the use of the AAVSO
Photometric All-Sky Survey (APASS), funded by the Robert Martin Ayers
Sciences Fund. Data products were also used from the Two Micron All
Sky Survey, which is a joint project of the University of
Massachusetts and the Infrared Processing and Analysis
Center/California Institute of Technology, funded NASA and the NSF.
The work has also made use of data from the European Space Agency
(ESA) mission {\it Gaia} (\url{https://www.cosmos.esa.int/gaia}),
processed by the {\it Gaia} Data Processing and Analysis Consortium
(DPAC,
\url{https://www.cosmos.esa.int/web/gaia/dpac/consortium}). Funding
for the DPAC has been provided by national institutions, in particular
the institutions participating in the {\it Gaia} Multilateral
Agreement. The computational resources used for this research include
the Smithsonian Institution's ``Hydra'' High Performance Cluster.

\end{acknowledgements}

\end{document}